\documentclass[journal=jctcce,manuscript=article,layout=onecolumn]{achemso}
\setkeys{acs}{articletitle = true}
\newcommand{\onlinecite}[1]{\citenum{#1}}
\usepackage[version=3]{mhchem} 
\usepackage{graphicx}
\usepackage{upgreek}
\usepackage{amsmath}
\usepackage{amssymb}
\usepackage{chemformula}
\usepackage{xcolor}
\usepackage{nicefrac}
\usepackage{booktabs}
\usepackage{comment}
\usepackage{mwe}
\usepackage{array}
\usepackage{syntonly}
\usepackage[utf8]{inputenc}
\usepackage{enumitem}

\usepackage{caption}
\usepackage{float}
\usepackage{subcaption}
\usepackage{graphicx}
\newcolumntype{m}{>{$} c <{$}}
\usepackage{color}

\usepackage[utf8]{inputenc}
\usepackage{verbatim}  
\usepackage{mathtools}
\usepackage[final]{pdfpages}
\usepackage{amsmath}
\usepackage{physics}
\usepackage{bm}
\usepackage[version=3]{mhchem} 
\usepackage{graphicx}
\usepackage{amsmath}
\usepackage{amssymb}
\usepackage{chemformula}
\usepackage{nicefrac}
\usepackage{booktabs}
\usepackage{mwe}
\usepackage{array}
\usepackage{syntonly}
\usepackage[utf8]{inputenc}
\usepackage{enumitem}
\usepackage{hyperref}
\hypersetup{
    colorlinks=true,
    linkcolor=blue,
    filecolor=blue,      
    urlcolor=blue,
    allcolors = blue
    }

\usepackage{caption}
\usepackage{float}
\usepackage{subcaption}
\usepackage{graphicx}
\usepackage{color}

\def\rv{{\bf r}}

\def\dd{\mathrm{d}}

\def\beq{\begin{equation}}
\def\eeq{\end{equation}}

\def\cc{\mathrm{c}}

\def\R{\mathbb{R}}

\DeclareMathOperator*{\argmin}{arg\,min}
\usepackage[version=3]{mhchem}

\newcommand{\di}{{\rm d}}
\newcommand{\ee}{{\rm e}}

\newcommand{\nulv}{{\bf 0}}

\newcommand{\bmath}{\begin{eqnarray}}
\newcommand{\emath}{\end{eqnarray}}

\newcommand{\rhoc}{{\rho_{\rm c}}}

\newcommand{\rhog}{{\Gamma}}
\newcommand{\gamg}{{\kappa}}

\newcommand{\rhoP}{\bar{\rho}}
\newcommand{\BP}{\widetilde{B}}

\title{Gradient expansions for the large-coupling strength limit of the M{\o}ller-Plesset adiabatic connection}

\author{Timothy J. Daas}
\affiliation
{Department of Chemistry \& Pharmaceutical Sciences and Amsterdam Institute of Molecular and Life Sciences (AIMMS), Faculty of Science, Vrije Universiteit, De Boelelaan 1083, 1081HV Amsterdam, The Netherlands}
\author{Derk P. Kooi}
\affiliation
{Department of Chemistry \& Pharmaceutical Sciences and Amsterdam Institute of Molecular and Life Sciences (AIMMS), Faculty of Science, Vrije Universiteit, De Boelelaan 1083, 1081HV Amsterdam, The Netherlands}
\author{Arthur J. A. F. Grooteman}
\affiliation
{Department of Chemistry \& Pharmaceutical Sciences and Amsterdam Institute of Molecular and Life Sciences (AIMMS), Faculty of Science, Vrije Universiteit, De Boelelaan 1083, 1081HV Amsterdam, The Netherlands}
\author{Michael Seidl}
\affiliation
{Department of Chemistry \& Pharmaceutical Sciences and Amsterdam Institute of Molecular and Life Sciences (AIMMS), Faculty of Science, Vrije Universiteit, De Boelelaan 1083, 1081HV Amsterdam, The Netherlands}
\author{Paola Gori-Giorgi}\email{p.gorigiorgi@vu.nl}
\affiliation{Department of Chemistry \& Pharmaceutical Sciences and Amsterdam Institute of Molecular and Life Sciences (AIMMS), Faculty of Science, Vrije Universiteit, De Boelelaan 1083, 1081HV Amsterdam, The Netherlands}
\begin{document}

\maketitle

\begin{abstract}
    The adiabatic connection that has as weak-interaction expansion the M{\o}ller-Plesset perturbation series has been recently shown to have a large coupling-strength expansion in terms of functionals of the Hartree-Fock density with a clear physical meaning. In this work we accurately evaluate these density functionals and we extract second-order gradient coefficients from the data for neutral atoms, following ideas similar to the ones used in the literature for exchange, with some modifications. These new gradient expansions will be the key ingredient for performing interpolations that have already been shown to reduce dramatically MP2 errors for large non-covalent complexes. As a byproduct, our investigation of neutral atoms with large number of electrons $N$ indicates that the second-order gradient expansion for exchange grows as $N\log(N)$ rather than as $N$ as often reported in the literature.
\end{abstract}

\section{Introduction}
Adiabatic connections (AC's) between an easy-to-solve hamiltonian and the physical, many-electron one, have always played a crucial role in building approximations in electronic structure theory. In density functional theory (DFT), the standard AC connects the Kohn-Sham (KS) hamiltonian with the physical one by turning on, via a parameter $\lambda$, the electron-electron interaction while keeping the one-electron density $\rho(\rv)$ fixed\cite{LanPer-SSC-75} (central column of tab~\ref{tab:ACs}). In this case, the series expansion of the correlation energy at small-coupling strengths ($\lambda\to 0$) is given by the G{\"o}rling-Levy (GL) perturbation theory. \cite{GorLev-PRB-93}  In the opposite limit of large-coupling strengths ($\lambda\to\infty$), the correlation energy is determined by the strictly-correlated-electrons (SCE) physics,\cite{Sei-PRA-99,SeiGorSav-PRA-07,Lew-CRM-18,CotFriKlu-ARMA-18} which yields the leading term. The next order is given by zero-point (ZP) oscillations\cite{GorVigSei-JCTC-09,GroKooGieSeiCohMorGor-JCTC-17,GroSeiGorGie-PRA-19,ColDimStra-arxiv-21} around the SCE manifold. A possible strategy to build approximations for the correlation energy is to interpolate between these two opposite limits, generalizing  to any non-uniform density\cite{SeiPerLev-PRA-99,SeiPerKur-PRL-00,GorVigSei-JCTC-09,LiuBur-PRA-09,VucGorDelFab-JPCL-18,GiaGorDelFab-JCP-18,Con-PRB-19} the idea that Wigner\cite{Wig-PR-34} used for jellium. The advantage of such an approach is that it is not biased towards the weakly-correlated regime. The lack of size-consistency of these interpolations can be easily corrected at zero computational cost.\cite{VucGorDelFab-JPCL-18} 

\begin{table}[]
\centering
\caption{The two adiabatic connections (AC's) considered in this work. Middle column: the standard density-fixed DFT AC that starts at $\lambda=0$ with the Kohn-Sham determinant. Right column: the AC that has the M{\o}ller-Plesset series as expansion for small coupling-strenghts $\lambda$ and starts at $\lambda=0$ with the Hartree-Fock Slater determinant.} \label{tab:ACs}
\begin{tabular}{l|l|l}
\multicolumn{1}{c|}{}       & \multicolumn{1}{c|}{} & \multicolumn{1}{c}{} \\
                        & DFT Adiabatic Connection & M{\o}ller-Plesset Adiabatic Connection \\\hline
$\hat{H}_{\lambda}$               &  $\hat{T}  + \hat{V}_{\rm ext} + \lambda\hat{V}_{\rm ee}+ \hat{V}_{\lambda}[\rho]$ & $\hat{T} + \hat{V}_{\rm ext} + \hat{V}^{\rm HF} + \lambda\left(\hat{V}_{\rm ee}-\hat{V}^{\rm HF}\right) $                    \\
                &                &      $\hat{V}^{\rm HF}=\hat{J}[\rho^{\rm HF}]-\hat{K}[\{\phi_{i}^{\rm HF}\}]$                \\
$\rho_{\lambda=0}$               &      $\rho$                &           $\rho^{\rm HF}$             \\
$\rho_{\lambda=1}$                &      $\rho$                  &          $\rho$              \\
$\rho_{\lambda}$                &       $\rho$                 &            $\rho_\lambda$            \\
$W_{c,\lambda}$                &  $\displaystyle \langle\Psi_{\lambda}|\hat{V}_{\rm ee}|\Psi_{\lambda}\rangle - \langle\Psi_{0}|\hat{V}_{\rm ee}|\Psi_{0}\rangle$                    &    $\displaystyle \langle\Psi_{\lambda}|\hat{V}_{\rm ee}-\hat{V}^{\rm HF}|\Psi_{\lambda}\rangle - \langle\Psi_{0}|\hat{V}_{\rm ee}-\hat{V}^{\rm HF}|\Psi_{0}\rangle$                   \\
$E_c$                         &   $\displaystyle \int_{0}^{1} W^{\rm DFT}_{c,\lambda} d\lambda$                 &     $\displaystyle \int_{0}^{1} W^{\rm HF}_{c,\lambda} d\lambda$                  \\
$W_{c,\lambda\rightarrow0}$    & $\displaystyle \sum_{n=2}^{\infty} n\,E_{c}^{{\rm GL}n}\lambda^{n-1}$             &  $\displaystyle \sum_{n=2}^\infty n\,E_{c}^{{\rm MP}n}\lambda^{n-1}$                   \\
$W_{c,\lambda\rightarrow\infty}$   &   $\displaystyle W^{\rm DFT}_{c,\infty}+ W^{\rm DFT}_{\frac{1}{2}}\lambda^{-1/2} +O\left(\lambda^{-5/4} \right)$                   &      $\displaystyle W^{\rm HF}_{c,\infty}+ W^{\rm HF}_{\frac{1}{2}}\lambda^{-1/2} + W^{\rm HF}_{\frac{3}{4}}\lambda^{-3/4} +O\left(\lambda^{-5/4} \right) $              
\end{tabular}
\end{table}

More recently,\cite{DaaFabDelGorVuc-JPCL-21} the same interpolation idea has been applied to the AC that has the M{\o}ller-Plesset (MP) series as perturbation expansion at small coupling strengths $\lambda$ (right panel of tab~\ref{tab:ACs}), connecting the Hartree-Fock (HF) hamiltonian with the physical one. The $\lambda\to\infty$ expansion of this MP AC is given by functionals of the Hartree-Fock density $\rho^{\rm HF}(\rv)$, with a clear physical meaning.\cite{SeiGiaVucFabGor-JCP-18,DaaGroVucMusKooSeiGieGor-JCP-20} 
The strong-coupling functionals of the DFT and the MP AC's are essentially electrostatic energies, whose exact evaluation for large particle numbers is demanding, but
while for the DFT AC there are rather accurate second-order gradient expansion approximations (GEA2)\cite{SeiPerKur-PRA-00,SeiGorSav-PRA-07,GorVigSei-JCTC-09} and, more recently, also generalised gradient approximations\cite{Con-PRB-19} (GGA), for the MP AC these approximations are not yet available. For this reason, in a very recent work\cite{DaaFabDelGorVuc-JPCL-21} the $\lambda\to\infty$ functionals of the MP AC have been modeled in an empirical way, starting from the GEA2 of the DFT ones. Quite remarkably, interpolations combined with this simple empirical model provide already very accurate results for non-covalent interactions (NCI), reducing the MP2 error by up to a factor 10 in the L7 dataset\cite{SedJanPitRezPulHob-JCTC-13}, without spoiling MP2 energies for the cases in which they are accurate.\cite{DaaFabDelGorVuc-JPCL-21} These interpolations work very well for diverse NCI's such as charge transfer and dipolar interactions, and they are able to correct MP2 both when it overbinds and when it underbinds, as they are able to take into account the change from concave to convex behavior of the MP AC.\cite{DaaFabDelGorVuc-JPCL-21} Their appealing feature is that they use 100\% of HF exchange and MP2 correlation energy, and it is the interpolation that decides for each system how much to correct with respect to MP2. This way, dispersion corrections are not needed at all to get accurate NCI's.\cite{DaaFabDelGorVuc-JPCL-21}

The purpose of this work is to derive the missing GEA2 for the strong-coupling functionals of the MP AC, in order to reduce empiricism and hopefully increase the accuracy of the interpolations along the MP AC. To this purpose, we use the ideas derived from the semiclassical limit of neutral atoms, which have been used in recent years in DFT for the analysis of the exchange and correlation functionals,\cite{PerConSagBur-PRL-06,EllBur-CJC-09,LeeConPerBur-JCP-09,CanCheKruBur-JCP-18,KapSanBhaWagChoBheYuTanBurLevPer-JCP-20,OkuBur-arxiv-21} yielding new approximations such as PBEsol.\cite{PerRuzCsoVydScuConZhoBur-PRL-08} As we shall see, the functionals we need to approximate allow us to probe more extensively these ideas, revealing several interesting features that could be used more generally to build DFT approximations.
We also notice that an additional term with respect to refs~\onlinecite{PerConSagBur-PRL-06,EllBur-CJC-09,PerRuzCsoVydScuConZhoBur-PRL-08} should be present in the second-order gradient expansion for exchange.

The paper is organised as follows. In sec~\ref{sec:largelambdaMP} we quickly review the large-coupling-strength functionals of the MP AC, discussing their physical meaning and the crucial differences with those of the DFT AC. Then in sec~\ref{sec:GEAEl} we focus on the gradient expansion of the leading term at large coupling strengths: we carry out an extensive analysis by filling more and more particles in a given density profile, and also by considering closed-shell neutral atoms and ions, up to the Bohr atom densities, which provide the limit of highly ionized atoms. We compute the functional in a numerically accurate way and determine a second-order gradient coefficient for the neutral-atoms case. We also discuss differences with the work of refs~\citenum{PerConSagBur-PRL-06,EllBur-CJC-09}, providing an analysis that should be relevant also for the exchange and correlation functionals of DFT. In sec~\ref{sec:GEAW1/2}, along similar lines, we extract the GEA2 coefficient for the next leading term of the MP AC large-coupling-strength expansion. The computational details are described in sec~\ref{sub:computationaldetails}, and the last sec~\ref{sec:conc} is devoted to conclusions and perspectives. More technical details, a curious behaviour of $N=2$ ions, and the discussion of an electrostatic model similar to the one used to derive the GEA2 coefficient of DFT are reported in the Appendix. Hartree atomic units will be used throughout this work.

\section{The large coupling-strength functionals of the M{\o}ller-Plesset AC}\label{sec:largelambdaMP}
\subsection{The M{\o}ller-Plesset AC}
To start, we need to introduce the M{\o}ller-Plesset Adiabatic Connection (MP AC), which has the following Hamiltonian,  
\begin{equation}\label{eq:HlambdaHF}
	\hat{H}_{\lambda}^{\rm HF}=\hat{T}+\hat{V}_{\rm ext}+\lambda \hat{V}_{ee}+
	\left(1-\lambda \right)
	\left(\hat{J}-\hat{K} \right),
\end{equation}
with $\hat{T}$ the kinetic energy, $\hat{V}_{ee}$ the electron-electron repulsion, and $\hat{V}_{\rm ext}$ the external potential due to the nuclei. The operators $\hat{J}=\hat{J}[\rho^{\rm HF}]$ and $\hat{K}=\hat{K}[\{\phi_i^{\rm HF}\}]$ are the standard Hartree-Fock (HF) Coulomb and exchange operators in terms of the HF density $\rho^{\rm HF}$ and the corresponding occupied orbitals $\phi_i^{\rm HF}$, respectively, which are determined in the initial HF calculation and do not depend on $\lambda$. Notice that in our notation $\hat{K}$ is positive definite. This Hamiltonian links the  Hartree-Fock system ($\lambda=0$) to the physical system ($\lambda=1$). The HF (or standard wavefunction theory) correlation energy, using the Hellmann-Feynman theorem, is given by
\beq \label{eq:ac}
E_{\rm c}^{\rm HF}= \int_0^1 W_{\rm c,\lambda}^{\rm HF} \dd \lambda,
\eeq
with $W_{\rm c,\lambda}^{\rm HF}$ the MP AC integrand,
\beq \label{eq:wchf}
W_{\cc, \lambda}^{\rm HF}=\langle  \Psi_{\lambda}| \hat{V}_{ee} - \hat{J} +\hat{K} | \Psi_{\lambda}  \rangle + U[\rho^{\rm HF}] + E_{x}^{\rm HF}[\{\phi^{\rm HF}_{i}\}],
\eeq
and $\Psi_{\lambda}$ the wave function that minimizes the expectation value of the hamiltonian of eq~\eqref{eq:HlambdaHF}. The last two terms, $U[\rho^{\rm HF}]$ and $E_{x}[\{\phi^{\rm HF}_{i}\}]$, are the Hartree energy and the HF exchange energy, respectively, whose sum gives minus the expectation value of $\hat{V}_{ee}-\hat{J}+\hat{K}$ on the HF Slater determinant (see right column of tab~\ref{tab:ACs}). The small-$\lambda$ expansion of $W_{\cc, \lambda}^{\rm HF}$ is the familiar MP perturbation series,
\begin{equation}\label{eq:WHFMP}
    W_{c,\lambda\rightarrow 0}^{\rm HF}=\sum_{n=2}^\infty n\,E^{{\rm MP}n}_{c}\,\lambda^{n-1}.
\end{equation}
\subsection{The $\lambda\to\infty$ expansion of the MP AC}
The large-$\lambda$ expansion of the MP AC has recently been uncovered\cite{SeiGiaVucFabGor-JCP-18,DaaGroVucMusKooSeiGieGor-JCP-20} for closed-shell systems as follows,
\begin{align}\label{eq:finalexp}
    W^{\rm HF}_{c,\lambda\rightarrow\infty} & = W^{\rm HF}_{c,\infty} + \frac{W^{\rm HF}_{\frac{1}{2}}}{\sqrt{\lambda}}+\frac{W^{\rm HF}_{\frac{3}{4}}}{\lambda^{\frac{3}{4}}}+\dots \\
	W^{\rm HF}_{c,\infty} & = E_{\mathrm{el}}[\rho^{\rm HF}]+E_x^{\rm HF}\label{eq:Wcinffinal}\\
	W^{\rm HF}_{\frac{1}{2}} & \approx  2.8687 \sum_{i=1}^N\left(\rho^{\rm HF}(\textbf{r}_i^{\rm min})\right)^{1/2}\label{eq:Wc1/2fin}\\
	W^{\rm HF}_{\frac{3}{4}} & \approx -1.272\sum_{\textbf{r}_{Z_k}}Z_k  \left(\rho^{\rm HF}(\textbf{r}_{Z_k})\right)^{1/4}.\label{eq:finalambda3/4}
\end{align}
The leading order, eq.~\eqref{eq:Wcinffinal}, contains the electrostatic-energy density functional $E_{\rm el}[\rho]$, which entails a classical electrostatic minimization,
\begin{align}\label{eq:El}
E_{\rm el}[\rho]&=\nonumber\min_{\Psi}\expval{\hat{V_{ee}}-\hat{J}[\rho]}{\Psi}+U[\rho]\\
                 &=\min_{\{\textbf{r}_{1}\ldots \textbf{r}_{N}\}}\left\{\frac{1}{2}\sum^{N}_{i\neq j=1}\frac{1}{\abs{\textbf{r}_{i}-\textbf{r}_{j}}}-\sum^{N}_{i=1}v_{\rm H}\left(\textbf{r}_{i};[\rho]\right)+U[\rho]\right\},
\end{align}
with
\bmath
v_{\rm H}\big(\rv;[\rho]\big)\;=\;\int\di\rv'\,\frac{\rho(\rv')}{|\rv-\rv'|},
\emath
and
\bmath
U[\rho]\;=\;\frac12\int\di\rv\int\di\rv'\,\frac{\rho(\rv)\,\rho(\rv')}{|\rv-\rv'|}.
\emath
The density functional $E_{\mathrm{el}}[\rho]$ can be understood as the total electrostatic energy of a distribution of $N$ negative point charges and continuous ``positive'' charges with density $\rho(\rv)$. In other words, the $\lambda\to\infty$ limit of the MP AC is a crystal of classical electrons bound by minus the Hartree potential generated by the HF density.\cite{SeiGiaVucFabGor-JCP-18,DaaGroVucMusKooSeiGieGor-JCP-20}
The resulting minimizing positions $\{\textbf{r}_{1}^{\rm min}\ldots \textbf{r}_{N}^{\rm min}\}$ in eq~\eqref{eq:El}, in turn, determine the next leading term,  for which eq~\eqref{eq:Wc1/2fin} provides a rigorous variational estimate for closed-shell systems.\cite{DaaGroVucMusKooSeiGieGor-JCP-20}  This term is given by zero-point oscillations around the minimizing positions enhanced by the exchange operator $\hat{K}$, which mixes in excited harmonic oscillator states.\cite{DaaGroVucMusKooSeiGieGor-JCP-20} 
Finally, the sum in eq~\eqref{eq:finalambda3/4} only runs over those minimizing positions of eq~\eqref{eq:El} that happen to be at a nucleus, and it is also a variational estimate.\cite{DaaGroVucMusKooSeiGieGor-JCP-20}
 These first three leading terms provide a rigorous framework to link MP perturbation theory with DFT, in terms of functionals of the HF density. In practice, we do not want to perform each time the classical minimization of eq~\eqref{eq:El}, which is known to have many local minima and whose cost increases rapidly with $N$. We rather wish to find good gradient expansion approximations for the first two terms in the expansion \eqref{eq:finalexp}. The third term, $W^{\rm HF}_{\frac{3}{4}}$ of eq~\eqref{eq:finalambda3/4}, could instead be approximated by making the assumption that in a large system there is one minimizing position at each nucleus, transforming it into a functional of the HF density at the nuclei.
\subsection{Comparison with the $\lambda\to\infty$ expansion of the DFT AC}
In a recent work where an interpolation for $W_{c,\lambda}^{\rm HF}$ between MP2 and the $\lambda\to\infty$ limit  has been built and tested, \cite{DaaFabDelGorVuc-JPCL-21} the functional $W_{c,\infty}^{\rm HF}$ of eq~\eqref{eq:Wcinffinal} has been approximated  in terms of the strong interaction limit of the DFT AC, using the following inequality\cite{SeiGiaVucFabGor-JCP-18}
\begin{equation}\label{eq:ineq}
    W^{\rm HF}_{c,\infty}\leq W^{\rm DFT}_{\infty}[\rho] + E_{x}^{\rm HF}.
\end{equation}
The DFT AC of the central column of tab~\ref{tab:ACs}, uses the hamiltonian,
\begin{equation}\label{eq:adiabDFT}
	\hat{H}_{\lambda}^{\rm DFT}=\hat{T}+\hat{V}_{\rm ext}+\lambda\,\hat{V}_{ee}+\hat{V}_{\lambda}[\rho],
\end{equation}
with $\hat{V}_{\lambda}[\rho]=\sum_{i=1}^N v_{\lambda}(\rv_i,[\rho])$ being the one-body potential that forces the density to be equal to the physical one for all values of $\lambda$. With this Hamiltonian, the KS exchange-correlation (XC) energy is given by
\begin{equation}\label{eq:CCIDFT}
	E_{xc}^{\rm DFT}[\rho]=\int_0^1 W_{\lambda}^{\rm DFT}[\rho]\,d\lambda,
\end{equation}
where the DFT coupling constant integrand is
\begin{equation}\label{eq:WlambdaDFT}
	W_{\lambda}^{\rm DFT}[\rho]\equiv\langle\Psi_{\lambda}^{\rm DFT}[\rho]|\hat{V}_{ee}|\Psi_{\lambda}^{\rm DFT}[\rho]\rangle - U[\rho],
\end{equation}
and $\Psi_{\lambda}^{\rm DFT}[\rho]$ the wave function that minimizes the expectation value of \eqref{eq:adiabDFT}.
Although the $\lambda\to\infty$ expansion of the DFT AC has a similar form as the MP AC one of eq~\eqref{eq:finalexp}, there are important differences between the two. The first one is the lack of the $\lambda^{-3/4}$ term in the DFT AC, which has the following large-coupling expansion\cite{SeiGorSav-PRA-07,GorVigSei-JCTC-09}
\begin{equation}
    W^{\rm DFT}_{\lambda\rightarrow\infty} = W^{\rm DFT}_{\infty} + \frac{W^{\rm DFT}_{\frac{1}{2}}}{\sqrt{\lambda}} + O\left(\lambda^{-5/4} \right).
\end{equation}
The reason why the MP AC can have a $\lambda^{-3/4}$ term is that in this case there is no constraint on the density, and the electrons thus localize around the minimizing positions $\{\textbf{r}_{1}^{\rm min}\ldots \textbf{r}_{N}^{\rm min}\}$. The density approaches asymptotically, as $\lambda\to\infty$, a sum of Dirac delta functions centered around these minimizing positions. If one of the $\textbf{r}_{i}^{\rm min}$ happens to be at a nucleus, the non-analyticity of the Coulomb nuclear attraction and of the cusp in the HF orbitals and density give rise to  this term.\cite{DaaGroVucMusKooSeiGieGor-JCP-20} In the DFT AC case, the density constraint enforces $\Psi^{\rm DFT}_\infty$ to be a superposition of infinitely many classical configurations,\cite{SeiGorSav-PRA-07} so  the one with an electron at a nucleus has infinitesimal weight. 

The inequality \eqref{eq:ineq} can be understood on simple physical terms: the functional $W_\infty^{\rm DFT}[\rho]$ can be reformulated as\cite{SeiPerKur-PRA-00}
\begin{equation}\label{eq:WinfDFTelec}
    W_\infty^{\rm DFT}[\rho]=\langle \Psi_\infty^{\rm DFT}[\rho]|\hat{V}_{ee}-\hat{J}[\rho]|\Psi_\infty^{\rm DFT}[\rho]\rangle +U[\rho],
\end{equation}
where we have simply used the fact that the expectation of $\hat{J}[\rho]$ on any wave function with density $\rho(\rv)$ is $2U[\rho]$. Then we can interpret\cite{SeiPerKur-PRA-00} $ W_\infty^{\rm DFT}[\rho]$ as the electrostatic energy of a system of classical electrons forced to have density $\rho$ immersed in a classical background of charge density $\rho$ of opposite sign. Notice that $\Psi_\infty^{\rm DFT}[\rho]$ {\em does not minimize} this electrostatic energy, but it is given by
\begin{equation}
    \Psi_\infty^{\rm DFT}[\rho]=\argmin_{\Psi\to\rho}\langle \Psi|\hat{V}_{ee}|\Psi\rangle.
\end{equation}
The functional $E_{\mathrm{el}}[\rho]$ of eq~\eqref{eq:El}, in contrast, is obtained by letting $\Psi$ {\em relax to its minimum} in eq~\eqref{eq:WinfDFTelec}, which directly implies
\begin{equation}\label{eq:ineq2}
    E_{\mathrm{el}}[\rho]\le W_\infty^{\rm DFT}[\rho].
\end{equation}
Adding $E_x^{\rm HF}$ to both sides of this inequality yields eq~\eqref{eq:ineq}.
\subsection{Semilocal functionals for the $\lambda\to\infty$ expansion of the DFT AC}
In Ref.~\onlinecite{DaaFabDelGorVuc-JPCL-21} parameters were added to both terms on the right-hand-side of equation \eqref{eq:ineq} to be fitted to the S22 dataset\cite{JurSpoCerHob-PCCP-06,TakHohMalMarSher-JCP-10}. 
This inequality was used due to the lack of approximations for $E_{\mathrm{el}}[\rho]$, but also to allow the functional to be more flexible to approximate the missing but very large second order term. Although the exact evaluation of $W^{\rm DFT}_{\infty}[\rho]$ is even more expensive than the one of $W^{\rm HF}_{\infty}[\rho]$, a cheap\cite{SeiPerKur-PRA-00} but accurate\cite{SeiGorSav-PRA-07,GorVigSei-JCTC-09} approximation called the Point Charge Plus Continuum (PC) model exists, which is a GEA2 functional
\begin{equation}\label{eq:PC}
W^{\rm PC}_{\infty}[\rho]=A^{\rm PC}\int \rho(\textbf{r})^{\frac{4}{3}}\,d\textbf{r}+ B^{\rm PC}\int \frac{\left|\nabla \rho(\textbf{r}) \right|^2}{\rho(\textbf{r})^{\frac{4}{3}}}\,d\textbf{r},
\end{equation}
with $A^{\rm PC}=\frac{-9}{10}\left(\frac{4\pi}{3}\right)^{\frac{1}{3}}$ and $B^{\rm PC}=\frac{3}{350}\left(\frac{3}{4\pi}\right)^{\frac{1}{3}}\approx 0.005317$. The PC model was built from the physical interpretation of  $W^{\rm DFT}_{\infty}[\rho]$ provided by eq~\eqref{eq:WinfDFTelec}: perfectly correlated electrons that need to minimise their interaction while giving the same density $\rho$ of the classical positive background will tend to {\em neutralize} the classical charge distribution $\rho$ (which is different than {\em minimize} the total electrostatic energy as in $E_{\mathrm{el}}[\rho]$). Along similar lines, by considering zero-point oscillations around the PC positions, a GEA2 functional for the second-order term, was constructed\cite{SeiPerKur-PRA-00}
\begin{equation}\label{eq:PCWinfP}
{W}^{\rm PC}_{\frac{1}{2}}[\rho]=C^{\rm PC}\int \rho(\textbf{r})^{3/2} {d\textbf{r}}+D^{\rm PC}\int \frac{|\nabla\rho(\textbf{r})|^{2}}{\rho(\textbf{r})^{7/6}} {d\textbf{r}}\,,
\end{equation}
with \(C^{\rm PC}=\frac{1}{2}(3\pi)^{1/2}\approx 1.535\) and\cite{GorVigSei-JCTC-09} \(D^{\rm PC}=-0.028957\), where $D^{\rm PC}$ is fixed to reproduce the Helium-atom exact result.\cite{GorVigSei-JCTC-09} In newer work by Constantin,\cite{Con-PRB-19} GGA functionals for both terms were derived to fix, among other things, the diverging asymptotics of the XC potentials.\cite{FabSmiGiaDaaDelGraGor-JCTC-19} However these GGA's have larger errors than the original PC model when compared with accurate SCE values for small atoms. Notice that, in contrast to the DFT AC (where self-consistent calculations should in principle carried out), we here do not need the functional derivatives of these quantities, as the MP AC is designed to directly give the HF correlation energy in a post self-consistent-field manner.

\section{Second order gradient expansion for $E_{\mathrm{el}}[\rho]$}\label{sec:GEAEl}
In this section we wish to derive a gradient expansion for $E_{\mathrm{el}}[\rho]$ of eq~\eqref{eq:El}. As detailed in appendix~\ref{sec:ElPC}, we cannot proceed along lines similar to the derivation of the PC model used for the DFT AC, because the charge distribution with the electrostatic energy $E_{\mathrm{el}}[\rho]$ cannot easily be divided into weakly interacting cells.
Moreover, we are only interested in $E_{\mathrm{el}}[\rho]$ for $\rho(\rv)$ that are HF densities of atoms and molecules. For this reason, we follow  the procedure used for the DFT exchange functional $E_x^{\rm DFT}[\rho]$ in refs.~\citenum{PerConSagBur-PRL-06,EllBur-CJC-09}, with some modifications. This procedure extracts the GEA2 coefficient from accurate data, and is very suitable because under uniform coordinate scaling at fixed particle number $N$,
\bmath \label{eq:unifcoordscal}
\rho_\gamma(\rv)\;=\;\gamma^3\rho(\gamma\rv)\qquad\Rightarrow\qquad\int\di\rv\,\rho_\gamma(\rv)=N\quad\text{(for all $\gamma>0$)},
\emath
our functional $E_{\rm el}[\rho]$  displays the same scaling behavior as exchange,
\bmath
E_{\rm el}[\rho_\gamma]\;=\;\gamma\,E_{\rm el}[\rho],
\label{scaEel}\emath
since $\;v_{\rm H}\big([\rho_\gamma],\rv\big)=\gamma\,v_{\rm H}\big([\rho],\gamma\rv\big)\;$ and $\;U[\rho_\gamma]=\gamma\,U[\rho]$.

In practice, we wish to find out whether for slowly varying densities $E_{\rm el}[\rho]$ is well approximated by a second-order gradient expansion
\bmath
E^{\rm GEA2}_{\rm el}[\rho] \;=\; \underbrace{A^{\rm HF}\int\di\rv\,\rho(\rv)^{4/3}}_{\displaystyle E^{\rm LDA}_{\rm el}[\rho]}
\;+\; B^{\rm HF}\int\di\rv\,\frac{|\nabla\rho(\rv)|^2}{\rho(\rv)^{4/3}}.
\label{EelGEA0}\emath
The powers $\rho(\rv)^{4/3}$ in the two terms of this expression are a necessary consequence of the exact scaling law of
eq~\eqref{scaEel}. Defining the usual reduced gradient $x$ of the density $\rho$,
\bmath\label{eq:xredgrad}
x\big([\rho],\rv\big)\;=\;\frac{\;|\nabla\rho(\rv)|\;}{\rho(\rv)^{4/3}},
\emath
which essentially gives the relative change of the density on the scale of the average interparticle distance,
eq~\eqref{EelGEA0} can also be written as
\bmath
E^{\rm GEA2}_{\rm el}[\rho]
\;=\;A^{\rm HF}\int\di\rv\,\rho(\rv)^{4/3}\left\{1\;+\;\frac{B^{\rm HF}}{A^{\rm HF}}\,x\big([\rho],\rv\big)^2\right\}.
\label{EelGEAs}\emath
The GEA2 expression should become more and more accurate as $x\to 0$, and our goal is to find the values of $A^{\rm HF}$ and $B^{\rm HF}$. As we will discuss later, while  $A^{\rm HF}$ is universal, the  coefficient $B^{\rm HF}$ seems to depend on how the slowly varying limit is approached, similarly to what happens with the DFT exchange functional.\cite{PerConSagBur-PRL-06,KapSanBhaWagChoBheYuTanBurLevPer-JCP-20}

\subsection{LDA coefficient $A^{\rm HF}$}
The uniform density limit $N\to\infty$ of a constant droplet density $\rho^{(r_s)}_N(\rv)=\frac3{4\pi\,r_s^3}\,\Theta(R_N-r)$ with radius $R_N=N^{1/3}r_s$, taken per particle
\bmath \label{eq:WignerCryst}
\lim_{N\to\infty}\frac{E_{\rm el}\big[\rho^{(r_s)}_N\big]}N
\;=\;\lim_{N\to\infty}\frac{A^{\rm HF}}N\int_{r\le R_N}\di\rv\,\left(\frac3{4\pi\,r_s^3}\right)^{4/3}
\;=\;A^{\rm HF}\cdot\left(\frac3{4\pi\,r_s^3}\right)^{1/3},
\emath
is equivalent to the jellium case\cite{LewLieSei-PRB-19} and has been analyzed already in Ref.~\onlinecite{DaaGroVucMusKooSeiGieGor-JCP-20}.
The result is the Wigner crystal energy per particle\cite{AlvBenEvaBer-PRB-21} $\;e_{\rm WC}(r_s)=-0.895 929 255\,\frac1{r_s}$, leading to
\bmath \label{eq:WignerCrystAHF}
A^{\rm HF}\;=\;-0.895 929 255\cdot\big({\textstyle \frac{4\pi}3}\big)^{1/3}\;=\;-1.44423075.
\emath
Notice that $A^{\rm HF}=A^{\rm DFT}$, where the latter is slightly different than the PC value $A^{\rm PC}$, which replaces $0.8959...$ with $0.9$. The fact that $A^{\rm DFT}$ is also exactly given by the Wigner crystal is proven rigorously in refs~\citenum{LewLieSei-PRB-19,CotPet-arxiv-17}.

\subsection{Particle-number scalings}\label{sub:TFscaling}
As discussed in refs.~\citenum{PerConSagBur-PRL-06,EllBur-CJC-09}, the slowly varying limit can be approached in different ways. An extended system with uniform density can be perturbed with a slowly-varying density distortion, but the resulting GEA2 coefficient might not be the one useful for chemistry.\cite{PerConSagBur-PRL-06} More generally,\cite{PerConSagBur-PRL-06,FabCon-PRA-13} for any functional that scales as eq~\eqref{scaEel}, we can reach the slowly varying limit by simply putting more and more electrons in a density profile $\rhoP(\rv)$
with $\int\di\rv\,\rhoP(\rv)=1$, by generating 
a discrete sequence of densities with increasing particle numbers $N=1,2,3,...$, using the scaling\cite{FabCon-PRA-13}
\bmath
\rhoP_{N,p}(\rv)\;=\;N^{3p+1}\,\rhoP\big(N^p\rv\big)\qquad\Rightarrow\qquad\int\di\rv\,\rhoP_{N,p}(\rv)=N.
\label{scaTF}\emath
With growing $N$, for all these densities the reduced gradient of eq~\eqref{eq:xredgrad} becomes increasingly weak,
\bmath
x\big([\rhoP_{N,p}],\rv\big)
\;=\;\frac{x\big([\rhoP],N^p\rv\big)}{N^{1/3}}\;\;\to\;\;0\qquad(N\to\infty),
\label{scaTF1}\emath
provided that
\bmath
\max_{\rv\in\R^3}\;x\big([\rhoP],\rv\big)\; {\rm is\; finite}.
\emath
Examples of relevant values of $p$ are
\begin{itemize}
    \item $p=\frac{1}{3}\;$: the Thomas-Fermi scaling of neutral atoms\cite{Lie-RMP-81,PerConSagBur-PRL-06,OkuBur-arxiv-21} $N=Z$;
    \item $p=-\frac{2}{3}\;$: the Thomas-Fermi scaling of the Bohr atoms\cite{HeiLie-PRA-95,StaScuPerTaoDav-PRA-04,OkuBur-arxiv-21,KapSanBhaWagChoBheYuTanBurLevPer-JCP-20};
    \item $p=0\;$: the scaling used in refs~\citenum{RasSeiGor-PRB-11,SeiVucGor-MP-16,VucLevGor-JCP-17} to analyze the Lieb-Oxford bound;
    \item $p=-\frac{1}{3}\;$: the scaling used in ref~\citenum{LewLieSei-PAA-20} to analyze the asymptotic exactness of the local density approximation.
\end{itemize}
For any density functional $G[\rho]$ that under the uniform coordinate scaling of eq~\eqref{eq:unifcoordscal} behaves as
\begin{equation}\label{eq:Gunif}
    G[\rho_\gamma]=\gamma^m \,G[\rho],
\end{equation}
for a fixed profile $\rhoP$ all the different choices of $p$ in eq~\eqref{scaTF} are equivalent and simply related to the case $p=0$,
\begin{equation}
    G[\rhoP_{N,p}]=N^{p\times m}\, G[\rhoP_{N,0}].
\end{equation}
For functionals that do not display a simple scaling behavior, like correlation in DFT, different values of $p$ lead to different interesting regimes, as discussed in refs~\citenum{PerConSagBur-PRL-06,FabCon-PRA-13}.

\begin{figure}
    \centering
    \includegraphics[width=0.6\textwidth]{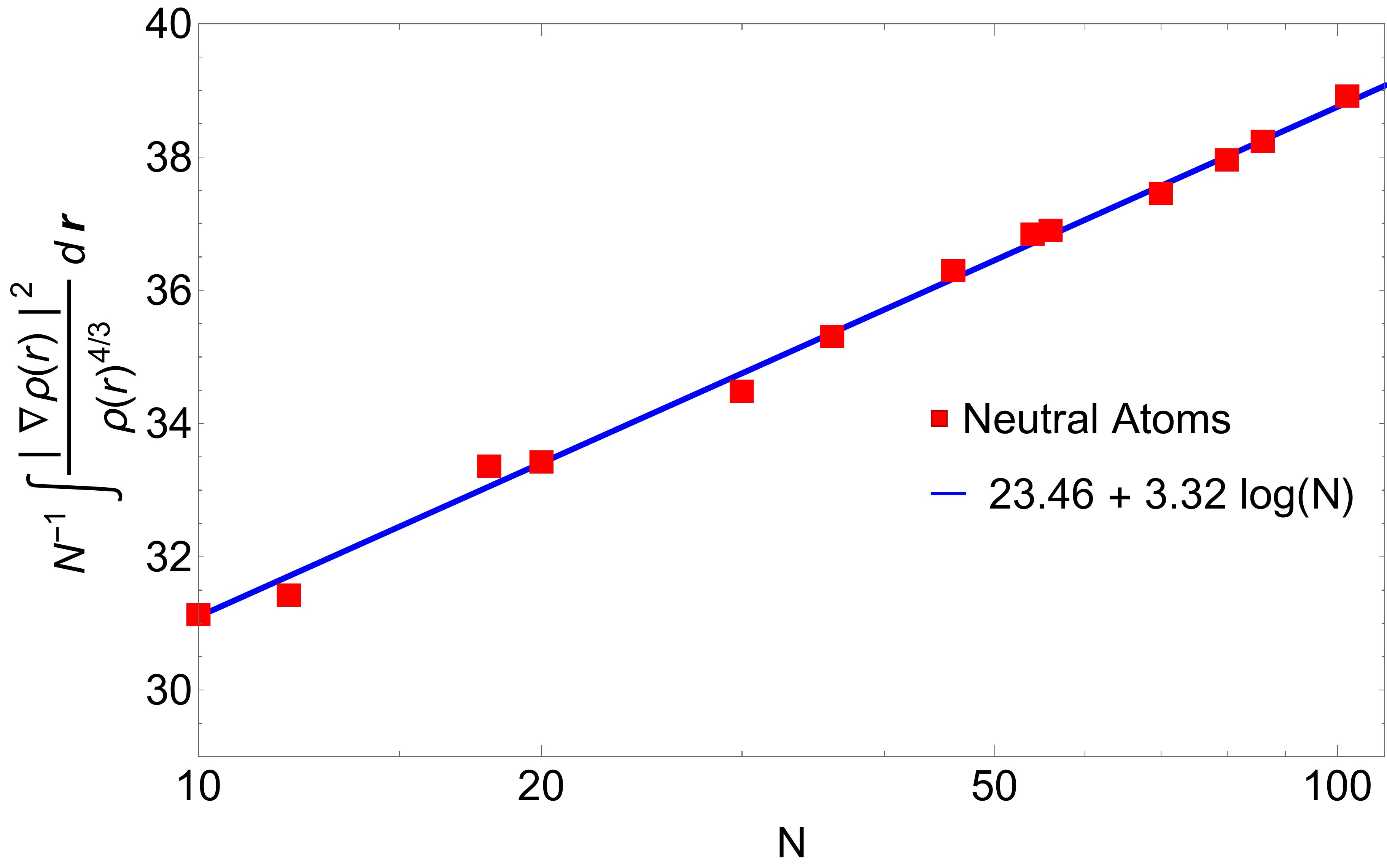}
    \caption{The GEA2 integral of eq~\eqref{eq:IGEA2} for the Hartree-Fock densities of neutral atoms, $I_{\rm GEA2}[\rho_{N=Z}]$, divided by the number of electrons $N$ (log scale on the $x$-axis). Numerical values (red dots) are compared with a logarithmic fit (blue line).} 
    \label{fig:NeutralLog}
\end{figure}

When $N$ grows, because of eq~\eqref{scaTF1}, we expect the gradient expansion of eq~\eqref{EelGEA0} to become more and more accurate. Then by inserting the density $\rhoP_{N,p}$ into eq~\eqref{EelGEA0} one gets the following large-$N$ expansion
\begin{equation}\label{eq:ElNexpa}
    E_{\mathrm{el}}[\rhoP_{N,p}]=N^{p+\frac{4}{3}}\,A^{\rm HF}\,I_{\rm LDA}[\rhoP]+N^{p+\frac{2}{3}}\,B^{\rm HF}\,I_{\rm GEA2}[\rhoP]+\dots\qquad  (N\to\infty),
\end{equation}
where
\begin{align}
    I_{\rm LDA}[\rhoP] & = \int \rhoP(\rv)^{4/3}\, d\rv, \label{eq:ILDA }\\
    I_{\rm GEA2}[\rhoP] & = \int \frac{|\nabla \rhoP(\rv)|^2}{ \rhoP(\rv)^{4/3}}\,d\rv. \label{eq:IGEA2}
\end{align}
Clearly, eq~\eqref{eq:ElNexpa} holds only as long as the integrals $I_{\rm LDA}[\rhoP]$ and $I_{\rm GEA2}[\rhoP]$ are finite for the given density profile $\rhoP$. 
In refs~\citenum{PerConSagBur-PRL-06,EllBur-CJC-09} the fact that the neutral atoms densities for large $N$ asymptotically satisfy the Thomas-Fermi (TF) scaling with $p=1/3$, 
\begin{equation}
    \rho_{N=Z}(\rv)\approx N^2 \rhoP_{\rm TFna}(N^{1/3}\rv)\qquad {\rm as}\;N\to\infty,
\end{equation}
with $\rhoP_{\rm TFna}(\rv)$ the TF profile (integrating to 1) for neutral atoms\cite{Lie-RMP-81,LeeConPerBur-JCP-09,CanCheKruBur-JCP-18,OkuBur-arxiv-21,KapSanBhaWagChoBheYuTanBurLevPer-JCP-20},
led to the conclusion that their exchange energy as a function of $N=Z$ should have, to leading orders, the large-$N$ expansion $a_x N^{5/3}+b_x N$. Extracting $b_x$ from exchange energies of neutral atoms allowed to fix the GEA2 coefficient for exchange in ref.~\citenum{EllBur-CJC-09}. 
However, while the GEA2 integral $I_{\rm GEA2}[\rho_{N=Z}]$ for neutral atoms is finite, the integral $I_{\rm GEA2}[\bar{\rho}_{\rm TFna}]$
for the asymptotic TF profile diverges (while $I_{\rm LDA}[\bar{\rho}_{\rm TFna}]$ is also finite). This does not automatically imply that 
$I_{\rm GEA2}[\rho_{N=Z}]$ should not increase linearly with $N$, as expected from eq~\eqref{eq:ElNexpa} with $p=\frac13$, since TF theory should not give 
exact information at this order. Nonetheless, we find numerical evidence (see fig~\ref{fig:NeutralLog}) that the GEA2 integral $I_{\rm GEA2}[\rho^{\rm HF}_{N=Z}]$ 
for Hartree-Fock densities of neutral atoms increases as $N\log(N)$ rather than as $N$. 

\begin{figure}[h]
    \centering
    \includegraphics[width=0.6\textwidth]{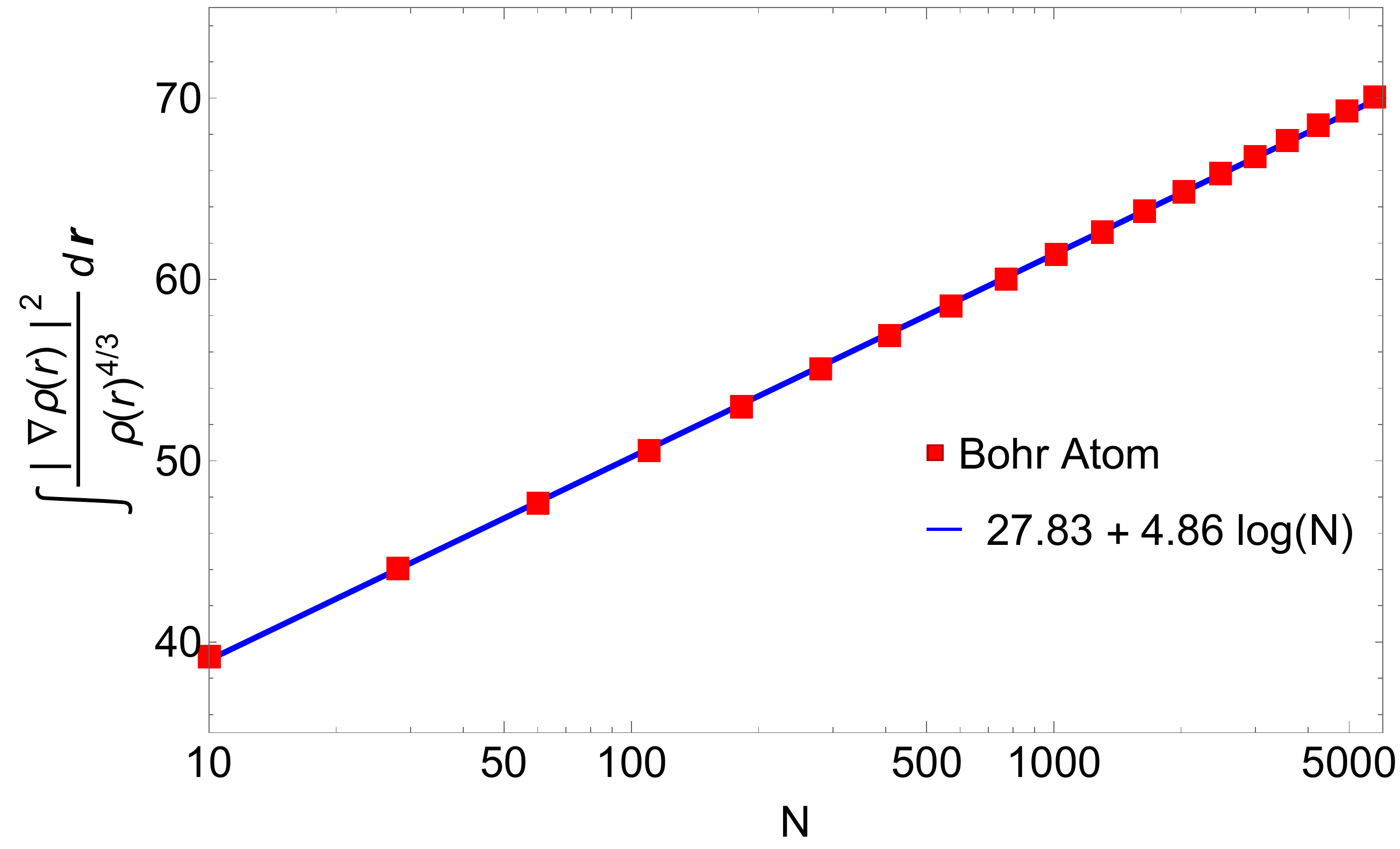}
    \caption{The GEA2 integral of eq~\eqref{eq:IGEA2} for the Bohr atom densities, $I_{\rm GEA2}[ \rho^{\rm Bohr}_N]$ (log scale on the $x$-axis). Numerical values (red dots) are compared with a logarithmic fit (blue line).} 
    \label{fig:BohrLog}
\end{figure}

A case for which it is even simpler to make a detailed numerical analysis of $I_{\rm GEA2}$ is the Bohr atoms,\cite{HeiLie-PRA-95,OkuBur-arxiv-21,KapSanBhaWagChoBheYuTanBurLevPer-JCP-20} which have densities constructed by occupying hydrogenic orbitals
\bmath
\rho^{\rm Bohr}_N(\rv)\;=\;2\,\sum_{n,\ell,m_\ell}\big|\psi_{n,\ell,m_\ell}(\rv)\big|^2,\label{eq:rhoBohr}
\emath
and can be thought\cite{HeiLie-PRA-95,KapSanBhaWagChoBheYuTanBurLevPer-JCP-20} as a limiting case for ions with $Z\gg N$. The latter have densities that, as $Z\to\infty$, approach those of the Bohr atom scaled as in eq~\eqref{eq:unifcoordscal} with $\gamma=Z$, 
\begin{equation}
\rho_{Z\gg N}(\rv)\approx Z^3\rho^{\rm Bohr}_N(Z\,\rv).
\end{equation}
As $N\to\infty$ the densities $\rho^{\rm Bohr}_N(\rv)$ of eq~\eqref{eq:rhoBohr} approach the Bohr atom TF profile\cite{HeiLie-PRA-95,OkuBur-arxiv-21,KapSanBhaWagChoBheYuTanBurLevPer-JCP-20} $\rhoP_{\rm TFBohr}$ with $p=-2/3$,
\begin{equation}
    \rho^{\rm Bohr}_N(\rv)\approx \frac{1}{N}\rhoP_{\rm TFBohr}(N^{-2/3}\rv) \qquad {\rm as}\;N\to\infty.
\end{equation}
Again, $I_{\rm LDA}[\rhoP_{\rm TFBohr}]$ is finite while $I_{\rm GEA2}[\rhoP_{\rm TFBohr}]$ diverges. If eq~\eqref{eq:ElNexpa} would hold with $p=-2/3$, $I_{\rm GEA2}[ \rho^{\rm Bohr}_N]$ should tend to a constant when $N\to\infty$. Instead, we clearly see (fig~\ref{fig:BohrLog}) that it grows as $\log(N)$. For this case everything is analytic and it is easy to reach very large $N$, evaluating the GEA2 integral to high accuracy.

A detailed derivation of the behaviour of $I_{\rm GEA2}[ \rho]$ as a function of $N$ for neutral atoms and for Bohr atoms, confirming the numerical evidence reported here, is also being carried out independently by Argaman et al.\cite{ArgRudCanBur-XXX-21}

\subsection{Extracting the GEA2 coefficient $B^{\rm HF}$}\label{sub:profiles}
\begin{figure}
    \centering
    \includegraphics[width=0.7\textwidth]{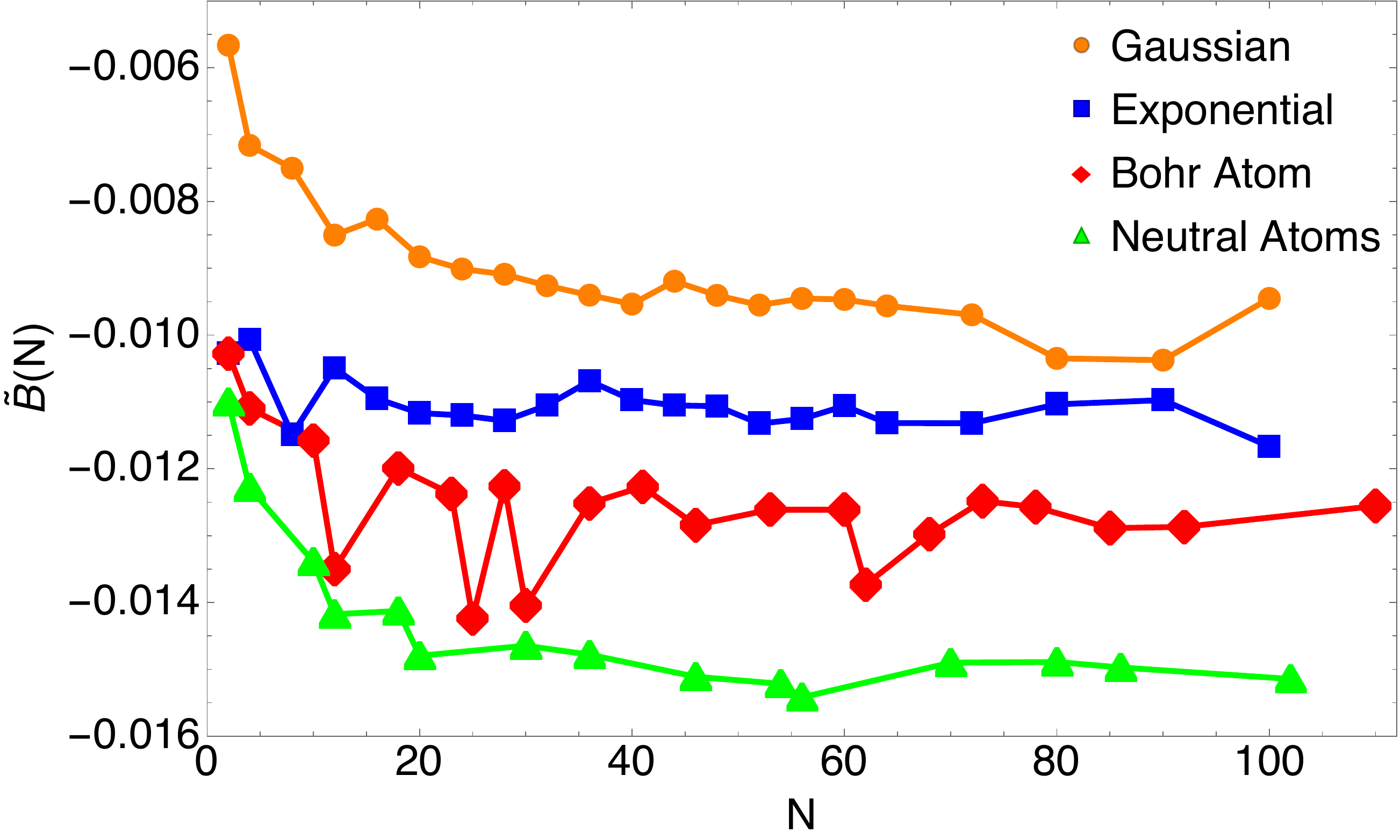}
    \caption{Equation \eqref{eq:Btilde} for the four cases described in sec~\ref{sub:profiles}.} 
    \label{fig:newBall}
\end{figure}

The analysis in the previous section suggests that extraction of the GEA2 coefficient should not be done by using values of $E_{\mathrm{el}}[\rho]$ as a function of $N$ and fitting coefficients from eq~\eqref{eq:ElNexpa}, as this seems to be safe only for a scaled known profile (as in eq~\eqref{scaTF}), but not for atomic densities. For this reason, we follow a route slightly different than the one used for exchange in ref~\citenum{EllBur-CJC-09}. Namely, we directly compute
\bmath \label{eq:Btilde}
\BP(N) = \frac{E_{\rm el}[\rhoP_N]\,-\,E^{\rm LDA}_{\rm el}[\rhoP_N]}
{\int\di\rv\,\frac{|\nabla\rhoP_N(\rv)|^2}{\rhoP_N(\rv)^{4/3}}}.
\emath
The idea is that if a GEA2 expansion for $E_{\mathrm{el}}[\rho]$ exists, we should observe that $\BP(N\to\infty)$ tends to a constant, which will be the sought $B^{\rm HF}$. However, such constant might not be the same for different profiles $\rhoP$ or when we use the neutral atoms or the Bohr atom densities. Indeed this seems to be the case: in fig~\ref{fig:newBall} we show for different particle numbers $N$
\begin{enumerate}
\item[(1)] Our numerical values $\BP(N)$ for the exponential profile
\bmath
\rhoP(r)\;=\;\frac1{8\pi}\,\ee^{-r}.
\emath
\item[(2)] Our numerical values $\BP(N)$ for the gaussian profile
\bmath
\rhoP(r)\;=\;\frac1{\pi^{3/2}}\,\ee^{-r^2}.
\emath
\item[(3)] Our numerical values $\BP(N)$ for the Hartree-Fock densities
$\rho^{\rm HF}_{N=Z}(\rv)$ of neutral atoms.
\item[(4)] Our numerical values $\BP(N)$ for the Bohr atom densities $\rho^{\rm Bohr}_N(\rv)$ of eq~\eqref{eq:rhoBohr}, including some cases in which we did not completely fill all the $\ell$ values for a given principal quantum number $n$. Notice that these latter cases cannot always be seen as the limit of highly ionized atoms, as degeneracy needs to be taken into account more carefully.
\end{enumerate}
The computational details behind the evaluation of $E_{\mathrm{el}}[\rho]$ for each case are described in sec~\ref{sub:computationaldetails}.
We see that these four sequences of data for $\BP(N)$ seem to approach four different limits as $N$ grows. Regarding the Bohr atoms, the cases for which the value of $\tilde{B}(N)$ suddenly drops to a value much closer to the one of neutral atoms are those in which we added an extra pair of $s$ electrons to a completely filled shell. For example, $N=12$ is obtained by adding $3s^2$ to the filled $n=2$ shell, and similarly for $N=30$ and $N=62$. The case $N=25$ is realised by filling the orbitals as in the Mn atom. From Fig. \ref{fig:newBall} we can conclude that there exists no unique GEA2 and that we should choose one of these $B^{\rm HF}$'s for our new GEA2 functional. As for the case of the exchange functional,\cite{PerConSagBur-PRL-06,EllBur-CJC-09} the most useful value for chemistry should be the one of neutral atoms.

\begin{figure}
    \centering
    \includegraphics[width=0.6\textwidth]{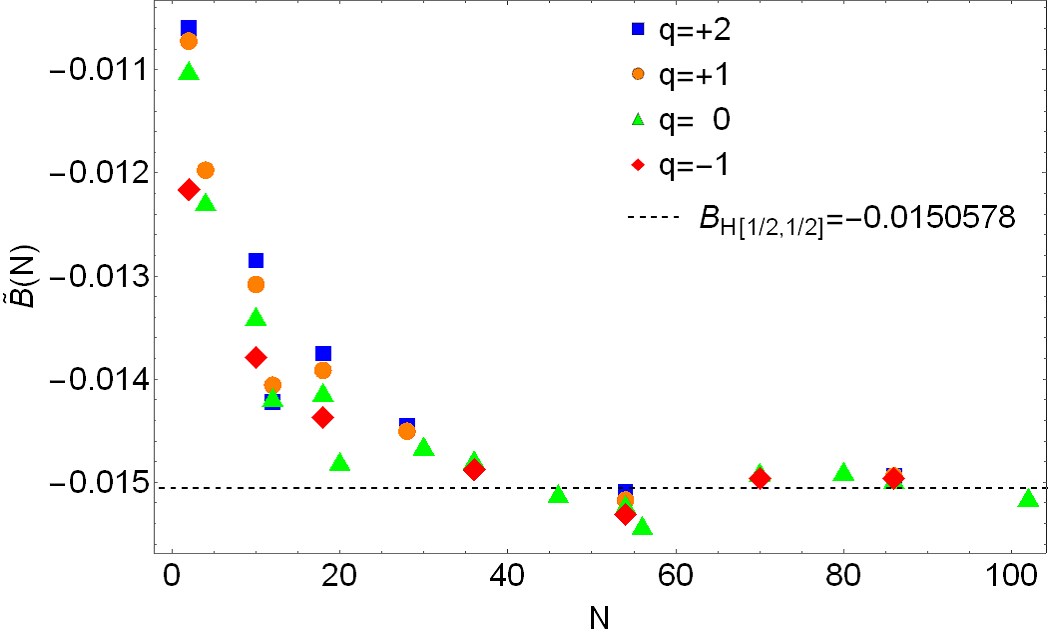}
    \caption{The value $ B^{\rm HF}_{\rm{H}[\frac{1}{2},\frac{1}{2}]}=-0.0150578$ that makes the GEA2 exact for the spin-unpolarised H atom accurately recovers the large $N$ limit of $\tilde{B}(N)$ of eq~\eqref{eq:Btilde} for closed-shell neutral atoms ($q=0$) and slightly charged closed shell ions, with $q=+1,+2$ and $-1$.} 
    \label{fig:BHs12}
\end{figure}

\begin{figure}
    \centering
    \includegraphics[width=0.6\textwidth]{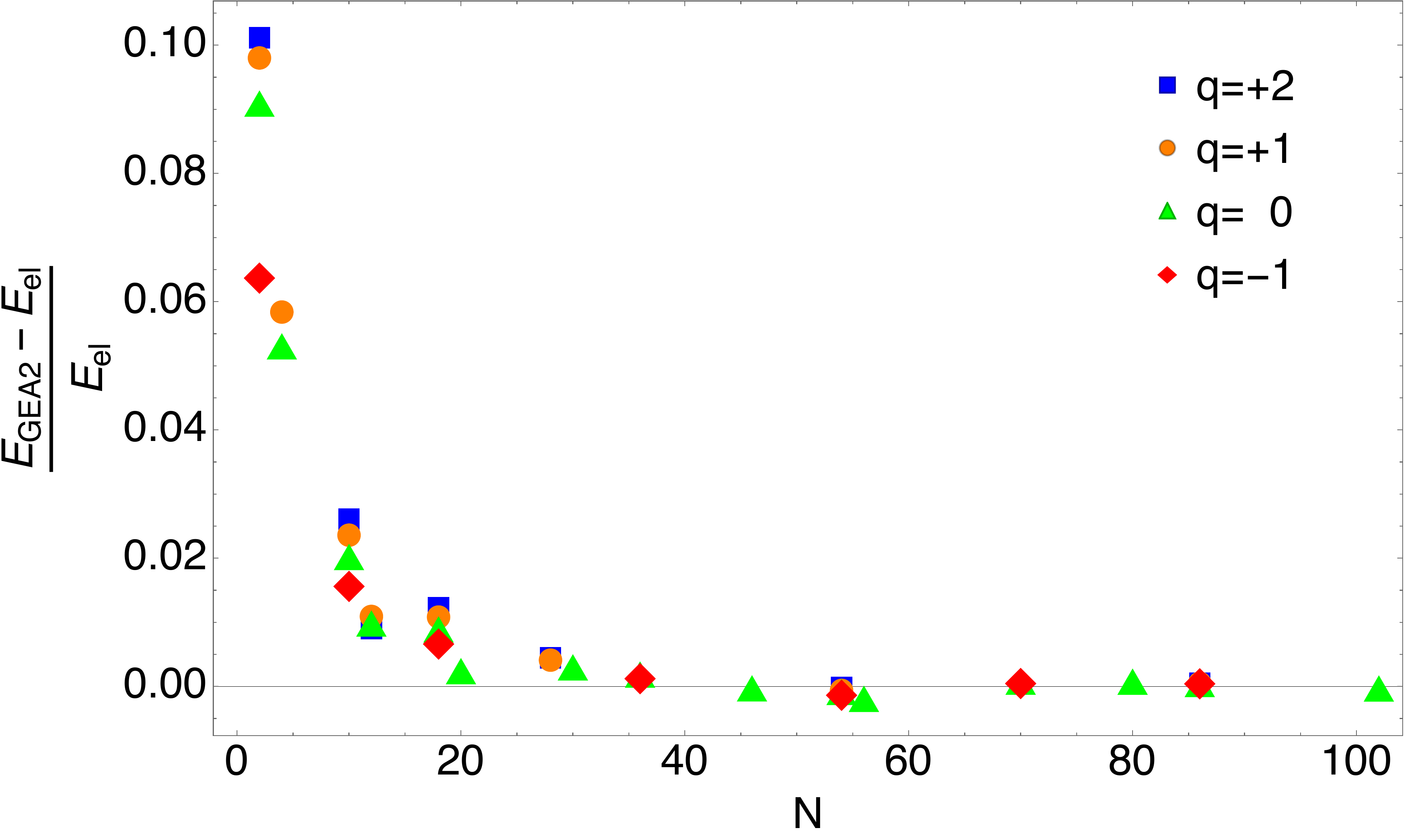}
    \caption{Relative error of the GEA2 expansion for the functional $E_{\mathrm{el}}[\rho]$ with $ B^{\rm HF}=B^{\rm HF}_{\rm{H}[\frac{1}{2},\frac{1}{2}]}=-0.0150578$ for closed shell neutral atoms $(q=0)$ and ions with $q=+1,+2$ and $-1$.} 
    \label{fig:errorGEA2}
\end{figure}

We noticed that if we fix $B^{\rm HF}$ to make the GEA2 exact for the spin-unpolarised H atom\cite{DaaGroVucMusKooSeiGieGor-JCP-20} (with $\frac{1}{2}$ spin-up and $\frac{1}{2}$ spin-down electrons\cite{CohMorYan-SCI-08}),
\begin{equation}\label{eq:BHhalfhalf}
    B^{\rm HF}_{\rm{H}[\frac{1}{2},\frac{1}{2}]}=-0.0150578,
\end{equation}
we recover the large $N$ limit of closed-shell neutral atoms and closed-shell ions with charges $+1$, $+2$ and $-1$ quite closely, as shown in fig~\ref{fig:BHs12}. We thus fix the GEA2 coefficient $B^{\rm HF}$ to this value, which seems to be as good as a fitted one, although we lack at this point a theoretical justification of why the H$[\frac{1}{2},\frac{1}{2}]$ should provide such a good number.
In fig~\ref{fig:errorGEA2} we show the relative error of the GEA2 expansion, which, as expected goes to zero for large neutral atoms and slightly charged ions.

\begin{figure}
    \centering
    \includegraphics[width=0.6\textwidth]{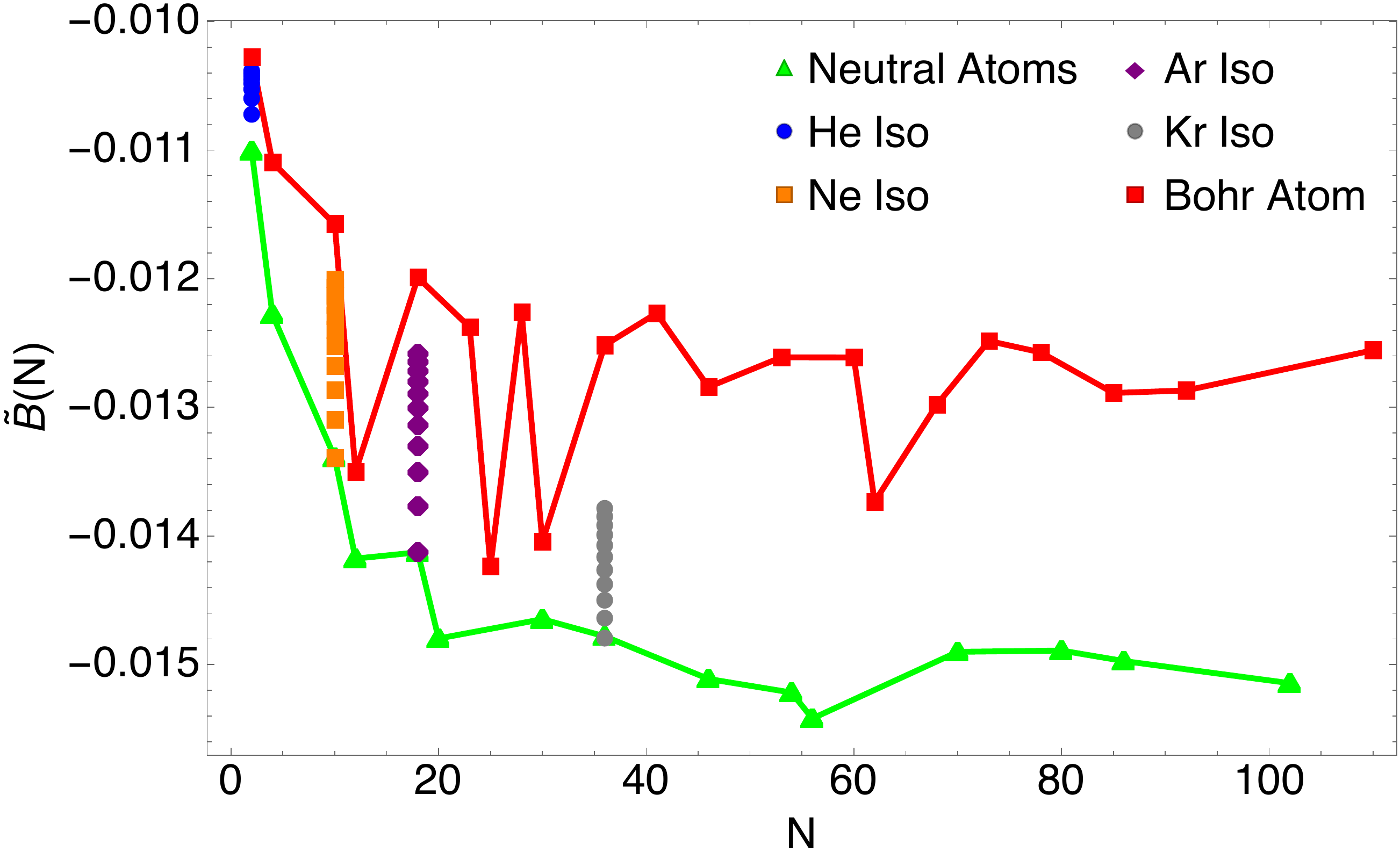}
    \caption{$\tilde{B}(N)$ of eq~\eqref{eq:Btilde} for neutral atoms, Bohr atoms and for selected noble-gas isoelectronic series. We see how  $\tilde{B}(N)$ goes from one limit to the other as the nuclear charge $Z$ is increased at fixed $N$. } 
    \label{fig:palette}
\end{figure}

We should however stress that the GEA2 with $B^{\rm HF}$ of eq~\eqref{eq:BHhalfhalf} misses the other\cite{KapSanBhaWagChoBheYuTanBurLevPer-JCP-20} slowly-varying limit of Bohr atoms with large $N$, which can be regarded as the limit\cite{KapSanBhaWagChoBheYuTanBurLevPer-JCP-20}  $Z\gg N \gg 1$. To better illustrate the issue, we show in fig~\ref{fig:palette} the values $\tilde{B}(N)$ only for the closed-shell neutral atoms, the Bohr atoms and for selected noble-gas isoelectronic series: we then see how  $\tilde{B}(N)$ goes from one limit to the other as the nuclear charge $Z$ is increased at fixed electron number $N$. 
An approximation able to cover this whole range of values could be maybe designed as a metaGGA, a route that will be investigated in future work.

\section{Second order gradient expansion for $W_{\frac{1}{2}}^{\rm HF}[\rho]$}\label{sec:GEAW1/2}
Once the minimization to obtain $E_{\mathrm{el}}[\rho]$ is performed, we automatically get the functional $W_{\frac{1}{2}}^{\rm HF}[\rho]$ of eq~\eqref{eq:Wc1/2fin} by evaluating the HF density in the minimizing positions $\rv_{i}^{\rm min}$. We should still stress that while the leading term of eq~\eqref{eq:Wcinffinal} is exact, eq~\eqref{eq:Wc1/2fin} is only a variational estimate valid for closed-shell systems within restricted HF.\cite{DaaGroVucMusKooSeiGieGor-JCP-20} Nonetheless we can repeat the analysis of the previous section to obtain a GEA2, which, due to the fact that $W_{\frac{1}{2}}^{\rm HF}[\rho]$ satisfies eq~\eqref{eq:Gunif} with $m=\frac{3}{2}$, must have the same form as the one for the DFT case of eq~\eqref{eq:PCWinfP},
\bmath
W_{\frac{1}{2}}^{\rm HF,GEA2}[\rho] \;=\; \underbrace{C^{\rm HF}\int\di\rv\,\rho(\rv)^{3/2}}_{\displaystyle W^{\rm HF,LDA}_{\frac{1}{2}}[\rho]}
\;+\; D^{\rm HF}\int\di\rv\,\frac{|\nabla\rho(\rv)|^2}{\rho(\rv)^{7/6}}
\label{eq:W1/2GEA2expr}\emath

\subsection{LDA coefficient $C^{\rm HF}$}
Within the variational expression of eq~\eqref{eq:Wc1/2fin}, the LDA coefficient $C^{\rm HF}$ is readily evaluated\cite{DaaGroVucMusKooSeiGieGor-JCP-20} and equal to $C^{\rm HF}=2.8687$. Notice that this is not the exact value for a uniform HF density, which should be evaluated by computing the normal modes around the bcc positions in the Wigner crystal and minimizing the total energy in the presence of the non-local operator $\hat{K}$, which will mix in excited modes. This analysis, using the techniques recently introduced by Alves et al.\cite{AlvBenEvaBer-PRB-21} is the object of a work in progress.
\subsection{Extraction of the GEA2 coefficient $D^{\rm HF}$}
\begin{figure}
    \centering
    \includegraphics[width=0.6\textwidth]{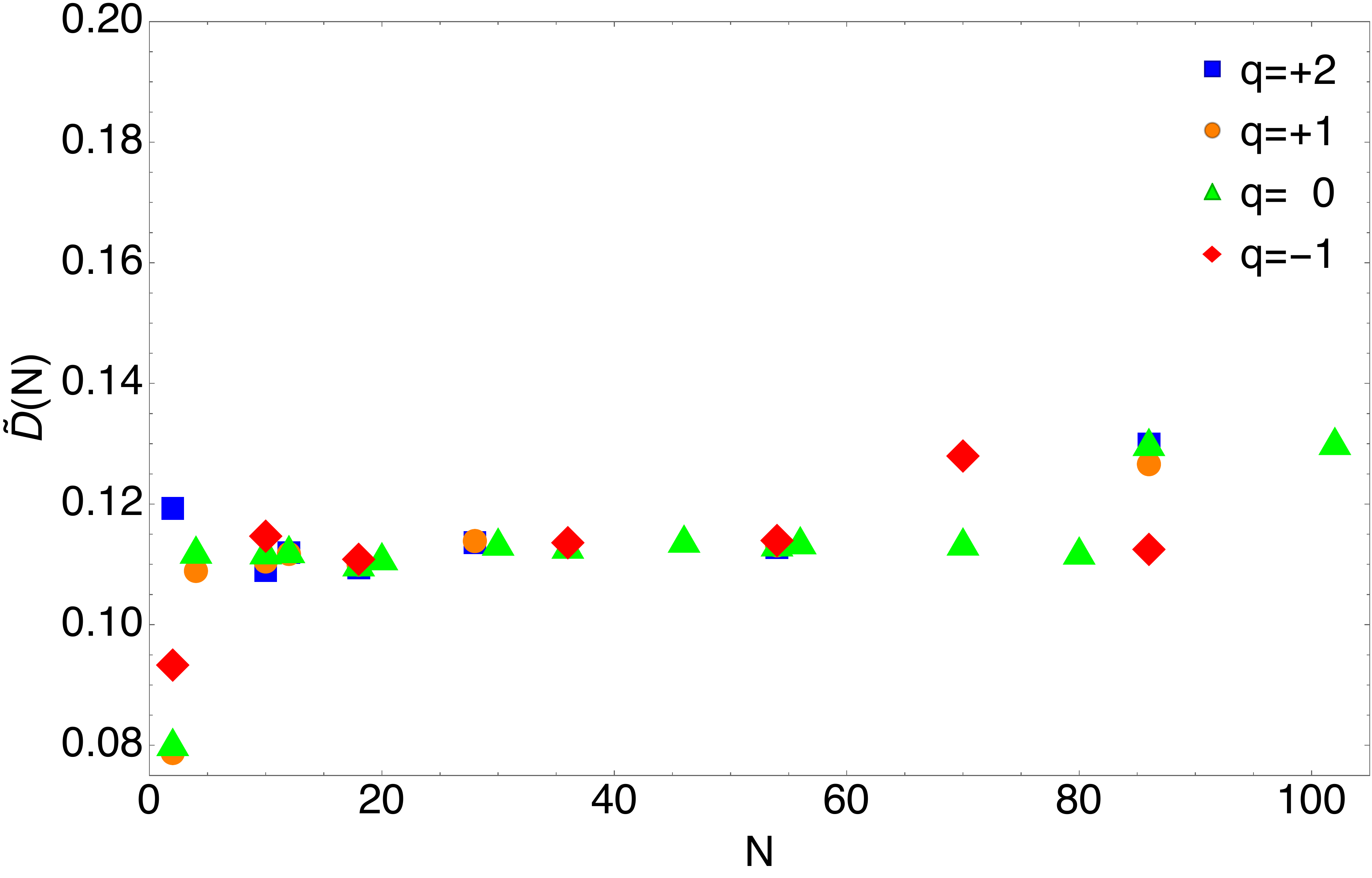}
    \caption{Values of $\tilde{D}(N)$ of eq~\eqref{eq:Dtilde} for closed-shell neutral atoms ($q=0$) and slightly charged closed shell ions, with $q=+1,+2$ and $-1$. }
    \label{fig:newD}
\end{figure}

We focus only on the relevant case of closed-shell neutral atoms and slightly charged ions, and, in analogy with eq~\eqref{eq:Btilde} we compute and analyse the function
\begin{equation}\label{eq:Dtilde}
    \tilde{D}(N)=\frac{W_{\frac{1}{2}}^{\rm HF}[\rho]-W^{\rm HF,LDA}_{\frac{1}{2}}[\rho]}{\int\di\rv\,\frac{|\nabla\rho(\rv)|^2}{\rho(\rv)^{7/6}}}.
\end{equation}
The results are shown in fig~\ref{fig:newD}, where we see that $\tilde{D}(N)$ gets rather flat already at $N\gtrsim 30$ around the value $\approx 0.11$. However, we also see a step to a slightly higher value, $\approx 0.13$, for the largest $N$. We don't know whether this step is really there or whether it is due to the numerical minimization being trapped in a local minimum. The issue is that as $N$ increases there are many local minima with very close values of $E_{\mathrm{el}}[\rho]$, which therefore remains rather insensitive if the true global minimum is not reached. The functional $W_{\frac{1}{2}}^{\rm HF}[\rho]$, however, depends on the minimizing configuration and it changes more from one local minimum to the other. We illustrate this in appendix~\ref{app:N2} for the case $N=2$, which undergoes a transition from a symmetric to an asymmetric minimum as the nuclear charge $Z$ varies from 2 to 1. From the data of fig~\ref{fig:newD} we can get a rough estimate $D^{\rm HF}\approx 0.12\pm 0.01$.

\section{Computational details}
\label{sub:computationaldetails}
To obtain reference values for $E_\mathrm{el}[\rho]$ for closed shell neutral atoms and slightly charged ions we first performed RHF calculations with PySCF 1.7.6,\cite{Sun-WIRCMS-18} with the basis sets specified in appendix~\ref{sub:basissets}. For a given set of positions $\{\rv_1, \rv_2, \dots \rv_N \}$ we calculated the value of $V_\mathrm{el}$, where

\begin{equation}
V_\mathrm{el}[\rho^\mathrm{HF}](\rv_1, \rv_2, \dots \rv_N ) =\left(\sum_{p < q} \frac{1}{|\mathbf{r}_p - \mathbf{r}_q|} - \sum_p v_\mathrm{H}[\rho^\mathrm{HF}](\mathbf{r}_p) \right) + U[\rho^\mathrm{HF}].
\end{equation}
We computed the value of $v_\mathrm{H}[\rho^\mathrm{HF}](\mathbf{r}_p)$ by contracting,
\begin{equation}
v_\mathrm{H}[\rho^\mathrm{HF}](\mathbf{r}_p) = \sum_{ij} \gamma_{ij}^{\mathrm{HF}} v^H_{ij}(\mathbf{r}_p), 
\end{equation}
where $\gamma^{\mathrm{HF}}$ is the Hartree-Fock 1-body Reduced Density Matrix (1-RDM) and the  matrix element is given by,

\begin{equation}
v^H_{ij}(\rv_p) = \int \di \rv \frac{\phi_i^*(\rv) \phi_j(\rv)}{|\rv-\rv_p|} \approx \int \di \rv \di \rv' \frac{\phi_i^*(\rv) \phi_j(\rv) G(\rv'-\rv_p)}{|\rv-\rv'|},
\end{equation}
where a very sharply peaked Gaussian $G$ was used to approximate the point-charge, which allows for a more efficient computation of the matrix elements using PySCF. To allow for minimization using a quasi-Newton method we also obtained the gradient of the Hartree potential,
\begin{align}
\nabla_{\rv_p} v_\mathrm{H}(\rv_p) &=\sum_{ij} \gamma^{\mathrm{HF}}_{ij} \left(\int \mathrm{d} \mathbf{r}' \frac{\left(\nabla_{\mathbf{r}'}\phi_i^*(\mathbf{r}') \right) \phi_j(\mathbf{r}')}{|\mathbf{r}'-\mathbf{r}_p|}+\int \mathrm{d} \mathbf{r}' \frac{\phi_i^*(\mathbf{r}') \left(\nabla_{\mathbf{r}'}\phi_j(\mathbf{r}')\right) }{|\mathbf{r}'-\mathbf{r}_p|} \right)\\
&\approx \sum_{ij}\gamma_{ij} \left(\int \mathrm{d} \mathbf{r} \mathrm{d} \mathbf{r}' \frac{\left(\nabla_{\mathbf{r}}\phi_i^*(\mathbf{r}) \right) \phi_j(\mathbf{r}) G_p(\mathbf{r}')}{|\mathbf{r}-\mathbf{r}'|}+\int \mathrm{d} \mathbf{r} \mathrm{d} \mathbf{r}' \frac{\phi_i^*(\mathbf{r}) \left(\nabla_{\mathbf{r}}\phi_j(\mathbf{r})\right) G_p(\mathbf{r}')}{|\mathbf{r}-\mathbf{r}'|} \right).
\end{align}
Then the total gradient is,
\begin{equation}
\nabla_{\rv_p} V_\mathrm{el}[\rho^\mathrm{HF}](\rv_1, \rv_2 \dots \rv_N) = - \frac{1}{2}\sum_{q \neq p} \frac{\rv_p-\rv_q}{|\rv_p-\rv_q|^3} - \nabla_{\rv_p} v_\mathrm{H}(\rv_p).
\end{equation}
Finally, $E_\mathrm{el}[\rho^\mathrm{HF}]$ was obtained by minimizing $V_\mathrm{el}[\rho^\mathrm{HF}]$ using the Broyden–Fletcher–Goldfarb–Shanno (BFGS) algorithm\cite{Bro-JAM-70,Fle-TCJ-70,Gol-MC-70,Sha-MC-70} as in the \texttt{scipy.optimize.minimize} function of \texttt{scipy}.\cite{2020SciPy-NMeth}

For selected cases, such as Ne and Ar, we have also double-checked the minimum by using Mathematica 12.3.1, experimenting with different minimizers. For the scaled densities and the Bohr atoms we have used both Mathematica and Python with the same \texttt{scipy.optimize.minimize} function used for the HF densities.

\section{Conclusions and Perspectives}\label{sec:conc}
We have built second-order gradient expansions for the functionals of the large-coupling-strength limit (see last line of the right column of tab~\ref{tab:ACs}) of the adiabatic connection that has the M{\o}ller-Plesset perturbation series as small-coupling-strength expansion, see eqs~\eqref{EelGEA0} and \eqref{eq:W1/2GEA2expr}. To this purpose, we have used ideas from the literature based on the semiclassical limit of neutral and highly ionized atoms.\cite{PerConSagBur-PRL-06,EllBur-CJC-09,OkuBur-arxiv-21} During our study we have also found numerical evidence (sec~\ref{sub:TFscaling} and figs~\ref{fig:NeutralLog}-\ref{fig:BohrLog}) which suggests that the way this semiclassical limit has been used to extract second-order gradient coefficients for exchange should be revised.\cite{PerConSagBur-PRL-06,PerRuzCsoVydScuConZhoBur-PRL-08,EllBur-CJC-09} 

In future work we will design and test new formulas for the adiabatic connection of the right-hand side of  tab~\ref{tab:ACs} that interpolate between MP2 and these new semilocal functionals at large-coupling, including the term proportional to $\lambda^{-3/4}$, which can be approximated as a functional of the HF density at the nuclei. Previous work\cite{DaaFabDelGorVuc-JPCL-21} showed that such functionals can be very accurate for non-covalent interactions, correcting the MP2 error for relatively large systems without using dispersion corrections. We will also analyse in the same way the functionals at strong coupling of the DFT AC (last line of the central column of tab~\ref{tab:ACs}), although in this case obtaining accurate results for large neutral atoms is numerically challenging. 

\section*{Acknowledgments}
We thank Kieron Burke and Nathan Argaman for confirming our numerical findings of figs~\ref{fig:NeutralLog}-\ref{fig:BohrLog}, for sharing a preliminary version of their independent work on the logarithmic contribution of the second-order gradient term for exchange,\cite{ArgRudCanBur-XXX-21} and for insightful discussions.
This work was funded by the Netherlands Organisation for Scientific Research under Vici grant 724.017.001. A.G. is grateful to the Vrije Universiteit for the opportunity to contribute to this paper using the University Research Fellowship.

\section*{Data Availability}
All the data for $E_{\rm el}[\rho]$ and $W^{\rm HF}_\frac{1}{2}[\rho]$, including the minimizing positions are available on \href{https://zenodo.org}{zenodo} at \href{https://doi.org/10.5281/zenodo.5734770}{10.5281/zenodo.5734771}.

\appendix
\section{Basis sets}
\label{sub:basissets}
For He-\ce{Zn^{2+}} we used aug-cc-pVQZ basissets from Ref. \onlinecite{Dun-JCP-89}. For the heavier atoms ranging from Zn to Xe we used Jorge augmented AQZP\cite{JorNetCamMac-JCP-09,PriAltDidGibWin-JCIM-19}, except for \ce{Br^{-}} where we used a standard aug-cc-pVQZ basisset. We used Jorge (augmented) ATZP for Cs to No, except for Ba and \ce{Ba^{2+}} for which we used Jorge TZP basisset. For the Helium iso-electronic series we used an aug-cc-pV6Z specifically designed for Helium~\cite{MouWilDun-MP-99}, whereas for the iso-electronic series of Neon and Argon we used a standard aug-cc-pV5Z basisset. For the Krypton iso-electronic series we used an aug-cc-pVQZ basisset instead.

\begin{figure}
    \centering
    \includegraphics[width=0.6\textwidth]{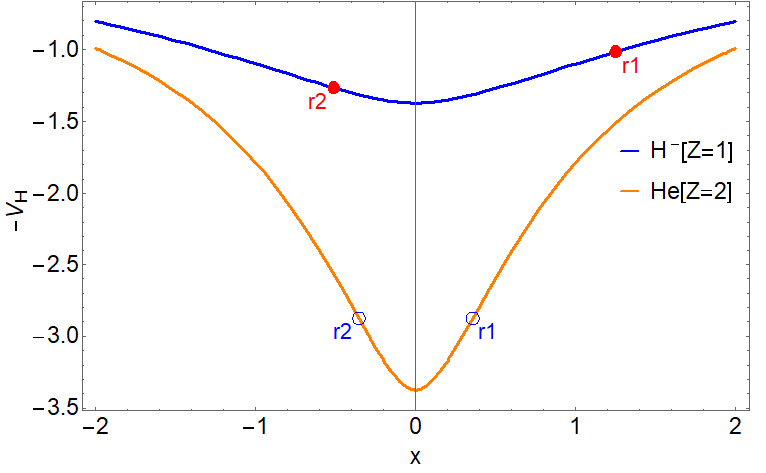}
    \caption{Minus the Hartree potential of the HF densities of He and H$^{-}$, together with the electronic positions that minimize $E_{\mathrm{el}}[\rho]$ in Hartree atomic units.}
    \label{fig:symmetrybreak}
\end{figure}

\begin{figure}
    \centering
    \includegraphics[width=0.6\textwidth]{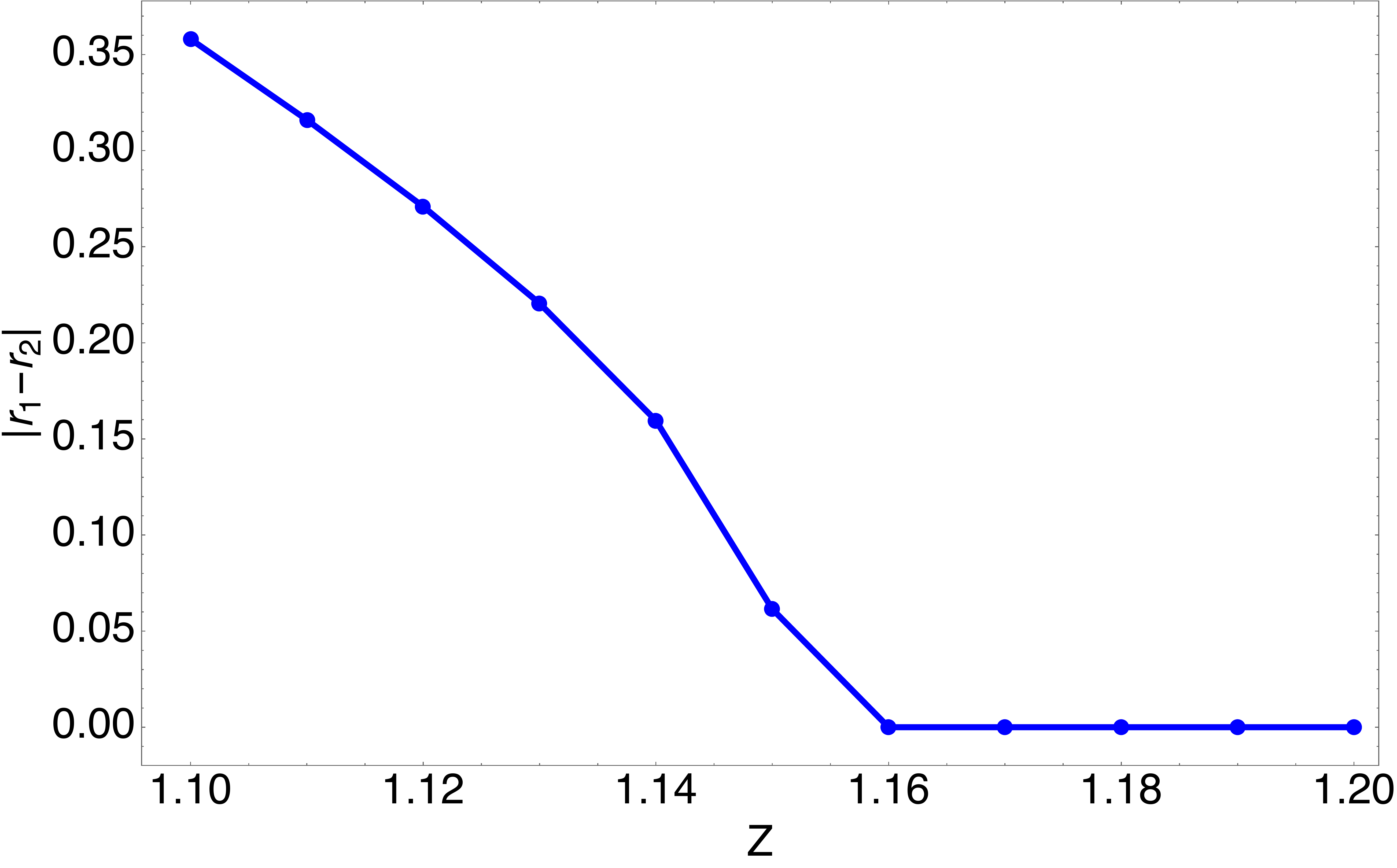}
    \caption{The difference between the distances from the nucleus of the two minimizing positions (Hartree atomic units) for $E_{\mathrm{el}}[\rho]$ for HF densities of ions with $N=2$ and varying nuclear charge $Z$.}
    \label{fig:deltar}
\end{figure}
\section{Symmetry breaking in $N=2$ ions}\label{app:N2}
Here we report a curiosity that we have observed about the minimizing positions of eq~\eqref{eq:El}. Naively, one would expect the minimizing positions for a $N=2$ atom to be symmetrically distributed on both sides of the well created by $-v_{\rm H}(r)$, which is the case for the Helium atom (orange curve) in fig~\ref{fig:symmetrybreak}, with minimizing positions in blue. However, when the nuclear charge $Z$ is decreased to 1 (H$^{-}$), the minimizing positions (red dots) are now asymmetrically distributed. In fig~\ref{fig:deltar} we show the difference between the distances from the nucleus of the two minimising positions as $Z$ varies: we see that the change from a symmetric to an asymmetric minimum happens at $Z=1.16$. In the case of H$^{-}$ a symmetric minimum is still present, but it is only a local one. In table~\ref{tab:Assym} we report the values of $E_{\mathrm{el}}[\rho]$ for H$^-$ for the asymmetric global minimum and the symmetric local one, showing that the difference is small (less than $0.1\%$), and any semilocal approximation for $E_{\mathrm{el}}[\rho]$ would have already larger errors than the difference between these two minima. However, the value of $W_{\frac{1}{2}}^{\rm HF}[\rho]$, also reported in the table, does change more significantly for the different minima (about 3\%), because it depends on the density at the minimizing positions directly, see eq~\eqref{eq:Wc1/2fin}. 

\begin{table}[]
\caption{The distances from the nucleus of the minimizing positions $r_1$ and $r_2$ (Hartree atomic units), $E_{\rm el}[\rho]$ and $W_{\frac{1}{2}}^{\rm HF}[\rho]$ of the symmetric local minimum and asymmetric global minimum of H$^{-}$.}
\begin{tabular}{l|ll}
\multicolumn{1}{c|}{} & \multicolumn{1}{c}{H$^{-}$[Sym]} & \multicolumn{1}{c}{H$^{-}$[Asym]} \\ \hline
$r_1$                    & 0.8477                          & 1.2515                           \\
$r_2$                    & 0.8477                          & 0.5116                           \\
$E_{\rm el}[\rho]$             & $-0.9219$                         & $-0.9228$                          \\
$W_{\frac{1}{2}}^{\rm HF}[\rho]$        & 1.4545                          & 1.5003                          
\end{tabular}
\label{tab:Assym}
\end{table}
\section{PC model for $E_{\rm el}[\rho]$ and its limitations}\label{sec:ElPC}
The PC model\cite{SeiPerKur-PRA-00} for $W^{\rm DFT}_{\infty}[\rho]$, eq \ref{eq:PC}, is built by starting from eq~\eqref{eq:WinfDFTelec} for a uniform density, in which the Wigner crystal of strictly-correlated electrons is approximated by having each electron surrounded by a sphere of background density exactly integrating to 1 (PC cell). Such an approximation amounts to replace with 0.9 the value of the bcc Madelung constant 0.8959.. of eq~\eqref{eq:WignerCrystAHF}. The GEA2 coefficient $B^{\rm PC}$ is then derived by applying a small gradient $\Gamma=|\nabla \rho(\rv)|$: by requiring that the PC cell plus its electron have exactly zero dipole moment, a new modified PC cell is constructed, whose center and size are slightly changed. This way, a given density can be thought as composed by cells (the PC cells plus their electron) that are weakly interacting, and the total energy can be obtained as a sum of the electrostatic energy of each cell (equal to the background-background interaction plus the electron-background interaction). 

In the case of $E_{\rm el}[\rho]$, instead, we want to minimize the total electrostatic energy. When the density is uniform, there is no other choice than creating the Wigner crystal bcc arrangement, which could be again approximated with PC spherical cells integrating to 1. Now we apply a small gradient and, in analogy with the DFT PC model, we focus on the energy of only one cell. The electron will move from the center of the sphere in the position that minimizes the electrostatic energy, i.e., the minimum of the Hartree potential of the PC cell with dipole. An estimate (with serious limitations discussed at the end of the derivation) of the resulting $B^{\rm HF\, PC}$  can then be easily computed as follows

We consider the charge density
\bmath
\rho(\rv)\;=\;\big(\rhoc\,+\,\rhog z\big)\cdot\Theta\big(R-|\rv|\big)
\emath
which is zero outside a sphere with radius $R$ (centered at the origin $\rv=\nulv$) and inside that sphere has a uniform gradient of magnitude $|\nabla\rho(\rv)|=\rhog$ in the $z$-direction, and
\bmath
\int\di\rv\,\rho(\rv)\;=\;\frac{4\pi}3R^3\,\rhoc\;=\;1.
\emath
The condition $\rho(\rv)\ge0$ implies
\bmath
\frac{\rhog R}{\rhoc}\;=\;\gamg\;\le\;1.
\emath
The Hartree potential (inside the sphere) and the Hartree energy are given by\cite{SeiPerKur-PRA-00}
\bmath
-v_{\rm H}\big([\rho],\rv\big)&=&-v_{\rm H}\big(r^2,z\big)\nonumber\\
&=&\left[\frac12\left(\frac{r^2}{R^2}\,-\,3\right)
\,+\,\frac{\gamg}{10}\,\frac{z}R\left(3\,\frac{r^2}{R^2}\,-\,5\right)\right]\frac{\left(\frac{4\pi}3R^3\rhoc\right)}R,\\
U[\rho]&=&\left[\frac35\,+\,\frac{\gamg^2}{35}\right]\frac{\left(\frac{4\pi}3R^3\rhoc\right)^2}R.
\emath
To find the minimizing position $\rv_{\rm min}\equiv(x_{\rm min},y_{\rm min},z_{\rm min})$ 
of the potential $-v_{\rm H}\big(r^2,z\big)$, we write $\bar{r}=\frac{r}R$, $\bar{z}=\frac{z}R$, 
\bmath
-v_{\rm H}\big(r^2,z\big)\;=\;\frac12\Big[\bar{r}^2\,+\,\frac35\,\gamg\,\bar{z}\,\bar{r}^2\,-\,3\,-\,\gamg\,\bar{z}\Big]\frac{1}R.
\label{vH_PC_HF}\emath
Setting here the partial derivatives (with respect to $\bar{x}$, $\bar{y}$, $\bar{z}$) zero, we obtain
\bmath
\bar{x}_{\rm min}&=&0,\nonumber\\
\bar{y}_{\rm min}&=&0,\nonumber\\
\bar{z}_{\rm min}&=&\frac{5}{9\gamg}\left[-1\,\pm\,\sqrt{1+\frac{9\gamg^2}5}\,\right]\qquad\text{(keep upper sign)}\nonumber\\
&=&\left[\frac{\gamg}2\,-\,\frac9{40}\gamg^3\,+\,O(\gamg^5)\right]\;\equiv\;\frac{z_{\rm min}}R.
\emath
The electron now sits at the position $\rv_{\rm min}=(0,0,z_{\rm min})$. Setting in Eq.~\eqref{vH_PC_HF} $\bar{r}=\bar{z}=\bar{z}_{\rm min}$ (and expanding around $\gamg=0$), then adding $U[\rho]$ to obtain $E_{\rm el}[\rho]$, yields
\bmath
-v_{\rm H}\big([\rho];0,0,z_{\rm min}\big)&=&\left[-\frac32\,-\,\frac{\gamg^2}8\,+\,O(\gamg^4)\right]\frac{1}{R},\\
E_{\rm el}^{\rm PC}[\rho]&=&\left[-\frac32\,-\,\frac{\gamg^2}8\right]\frac{1}{R}
            \;+\;\left[\frac35\,+\,\frac{\gamg^2}{35}\right]\frac{1}{R}\;+\;O(\gamg^4)\nonumber\\
&=&\left[-\frac9{10}\,-\,\frac{27}{280}\,\gamg^2\right]\frac1R\;+\;O(\gamg^4). \label{eq:BHFPC}
\emath
Replacing the cell radius $R$ with the local Seitz radius $r_s(\rv)$, this result should be compared with the PC DFT one in Eq.~(22) of ref~\citenum{SeiPerKur-PRA-00},
\bmath
E^{\rm PC\,DFT}_{\rm cell }\big([\rho];\rv\big)
\;=\;\left[-\frac9{10}\,+\,\frac{3}{350}\,\gamg^2\right]\frac1{r_{\rm s}(\rv)}\;+\;O(\gamg^4).
\emath
We see that this simple model correctly captures the sign of $B^{\rm HF}$, which is negative, contrary to the positive sign for the DFT case. The order of magnitude is also correct, as eq~\eqref{eq:BHFPC} yields $B^{\rm HF\,PC}=-\frac{27}{280}(\frac{4\pi}{3})^{-1/3}=-0.0598195$. However, comparison with the results of sec~\ref{sec:GEAEl} shows that this $B^{\rm HF\,PC}$ is too large in magnitude by almost a factor 4. We believe that the reasons for this discrepancy are: (i) the fact that the PC cells with their electron have now a dipole moment invalidates the hypothesis that the energy can be computed as a sum of the energies of weakly interacting cells. The total energy is probably raised due to the cell-cell interaction that might also alter the position of the electron and the size of the PC cell; (ii) similarly to the exchange functional of DFT,\cite{PerConSagBur-PRL-06} the GEA2 coefficient obtained from a uniform system weakly perturbed is probably not equal to the result obtained for neutral atom densities.

\bibliography{bib_clean.bib}

\end{document}